\documentclass[12pt]{article}
\usepackage{graphics}
\usepackage{amssymb}
\usepackage{amsmath}

\newcommand{\rf}[1]{(\ref{#1})}
\newcommand{\beq}{\begin{equation}}
\newcommand{\eeq}{\end{equation}}
\newcommand{\bea}{\begin{eqnarray}}
\newcommand{\eea}{\end{eqnarray}}

\newcommand{\e}{\mbox{e}}
\renewcommand{\d}{\mbox{d}}

\newcommand{\m}{\mu}

\newcommand{\sg}{\sigma}
\newcommand{\ra}{\rangle}

\newcommand{\prt}{\partial}
\newcommand{\mi}{\!-\!}
\newcommand{\equ}{\!=\!}
\newcommand{\pl}{\!+\!}

\newcommand{\tP}{{\tilde{P}}}
\newcommand{\tG}{{\tilde{G}}}

\newcommand{\hn}{{\hat{n}}}

\newcommand{\bx}{{\bar{x}}}
\newcommand{\bt}{{\bar{t}}}

\newcommand{\sm}{\sqrt{\m}}

\begin{document}

\begin{center}
\vspace{24pt} { \large \bf World-sheet dynamics of ZZ branes}

\vspace{30pt}

{\sl J. Ambj\o rn}$\,^{a,c}$
and
{\sl J. A. Gesser}$\,^{a}$.

\vspace{24pt}
{\footnotesize

$^a$~The Niels Bohr Institute, Copenhagen University\\
Blegdamsvej 17, DK-2100 Copenhagen \O , Denmark.\\
{ email: ambjorn@nbi.dk, gesser@nbi.dk}\\

\vspace{10pt}

$^c$~Institute for Theoretical Physics, Utrecht University, \\
Leuvenlaan 4, NL-3584 CE Utrecht, The Netherlands.\\

\vspace{10pt}

}
\vspace{48pt}

\end{center}

%\addtolength{\baselineskip}{0.20\baselineskip}
%\vspace{2cm}

\begin{center}
{\bf Abstract}
\end{center}

We show how non-compact space-time (ZZ branes) emerges as a limit of
compact space-time (FZZT branes) for specific ratios between the
square of the boundary cosmological constant and the bulk
cosmological constant in  the \mbox{$(2,2m \mi 1)$} minimal model
coupled to two-dimensional quantum gravity.

\vspace{12pt}
\noindent

%\vfill

\newpage

\section{Introduction}

Non-critical string theory serves as a good laboratory for the study
of non-perturbative effects in string theory. Since non-critical
string theory can also be viewed as two-dimensional gravity coupled
to matter, it also serves as a model for quantum gravity.

Most recently
%it was shown that
the dynamics of D-branes was studied in non-critical string theory
with $c < 1$ \cite{martinec,sei1,sei2}, where $c$ denotes the
central charge of the conformal field theory coupled to quantum
gravity. The starting point for this new development was the work of
Zamolodchikov and Zamolodchikov (ZZ) \cite{zz}, who from a purely 2d
gravity point of view asked if it is possible to quantize
non-compact 2d Euclidean geometries. The consistency conditions
imposed on  the quantization of the Lobachevskiy-plane (the
pseudo-sphere) led to the discovery of the boundary conditions at
infinity, which later, in \cite{martinec,sei1,sei2}, were
reinterpreted as branes (the so-called ZZ branes) in the context of
non-critical string theory.

There exists an intriguing relation between the ZZ branes of the
non-compact pseudo-sphere and the more conventional boundary
conditions on compact geometries analyzed by Fateev, Zamolodchikov
and Zamolodchikov and by Techner (the FZZT branes) \cite{fzzt}, a
relation first realized by Martinec \cite{martinec} and by Seiberg
and Shih \cite{sei1}, see also \cite{martinec2}. It is the purpose
of this article to provide an interpretation of this relation from
the viewpoint of the world-sheet theory, i.e.\ we ask what is the
physics behind the transition from the compact world-sheet
geometries characterizing the FZZT branes to the non-compact
geometries characterizing the ZZ branes. This world-sheet
perspective is interesting from a 2d quantum gravity point of view.

%\section{The disk and the cylinder amplitude}
\section{From compact to non-compact geometry}

The general disk and cylinder amplitudes in non-critical string
theory were first calculated using matrix model techniques. In order
to compare with continuum calculations, performed in the context of
Liouville theory, it is necessary to work in the so-called conformal
background \cite{staudacher}. In the following we will, for
simplicity, concentrate on the disk and the cylinder amplitudes in
the $(2,2m \mi 1)$ minimal conformal field theories coupled to 2d
quantum gravity, also called $(2,2m \mi 1)$ non-critical string
theories. The disk amplitude, calculated from the one-matrix model,
is \cite{kazakov}: \beq\label{1} w(x) =
(-1)^{m}\hat{P}_m(x,\sm)\sqrt{x+\sm} =
(-1)^{m}\left(\sm\right)^{(2m-1)/2} P_m(t)\sqrt{t+1},
%~~~t=x/\sm,
\eeq where $t=x/\sm$ and where the polynomial $P_m(t)$ is of degree
$m\mi 1$. In the conformal background it is determined by
\cite{staudacher} \beq\label{2} P^2_m(t)\;(t+1) =
2^{2-2m}(T_{2m-1}(t)+1) \eeq $T_p(t)$ being the first kind of
Chebyshev polynomial of degree $p$. In eq.\ \rf{1} $x$ denotes the
boundary cosmological coupling constant and $\m$ the bulk
cosmological coupling constant, the theory viewed as 2d quantum
gravity coupled to the $(2,2m \mi 1)$ minimal CFT. The zeros of
$P_m(t)$ are all located on the real axis between $\mi 1$ and $1$
and more explicitly we can write: \beq\label{2a} P_m(t) =
\prod_{n=1}^{m-1} (t-t_n),~~~~~ t_n =
-\cos\left(\frac{2n\pi}{2m-1}\right),~~~~1\le n \le m-1. \eeq The
zeros of $P_m(t)$ can be associated with the $m\mi 1$ principal ZZ
branes in the notation of \cite{sei1}.

%In the case of pure 2d gravity, i.e.\ $m\equ 2$, the zero of
%$\hat{P}_2(x)$ is related to a divergent geodesic distance on the
%world-sheet \cite{aagk} due to the relation \beq\label{3}
%d=\int_{\bx(d)}^x \frac{\d x'}{w(x')}, \eeq where $d$ is the
%geodesic distance and $\bx(d)$ the so-called running boundary
%cosmological coupling constant. One observes that when $d \to \infty
%$ the running boundary coupling constant $\bx(d)$ converges to  the
%zero of $\hat{P}_2(x)$.
The so-called loop-loop propagator $G_\m(l_1,l_2;d)$
\cite{kawai1,kawai2,gk,aw_pq} is well suited to show the transition from
a compact to a non-compact space.
%when $\bx$ approaches a zero
%of $P_2(x)$
It describes the amplitude of an ``exit'' loop of length $l_2$ to be
separated a distance $d$ from  an ``entrance'' loop of length $l_1$
(the entrance loop conventionally assumed to have one marked point).
By Laplace transformation one can introduce the boundary
cosmological constants $x,y$ of the entrance and the exit loops,
i.e. FZZT-branes on both boundaries: \beq\label{4} G_\m(x,y;d) =
\int_0^\infty \int_0^\infty \d l_1\d l_2 \;
\e^{-l_1x}\e^{-l_2y}\;G_\m(l_1,l_2;d). \eeq It can be shown
\cite{kawai1,kawai2,gk} that $G_\m(x,y;d)$ satisfies the following
equation: \beq\label{5} \frac{\prt }{\prt d}\; G_\m (x,y;d) = -
\frac{\prt }{\prt x} \; w(x) G_\m (x,y;d), \eeq with the following
solution: \beq\label{6} G_\m (x,y;d) = \frac{w(\bx(d))}{w(x)} \;
\frac{1}{\bx(d)+y}, \eeq where the so-called running boundary
coupling constant $\bx (d)$ is the solution of the characteristic
equation corresponding to \rf{5}, i.e.
%\ defined by eq.\ \rf{3}.
\beq\label{3} d=\int_{\bx(d)}^x \frac{\d x'}{w(x')} \eeq

While $G_\m(x,y;d)$ is not a much studied object in 2d quantum
gravity, it is actually a kind of fundamental building block: In
pure quantum gravity knowing $G_\m(x,y;d)$ allows one to calculate
the cylinder amplitude $C(x,y;\m)$ and in principle all higher loop
functions $C(x_1,\ldots,x_n;\m)$) \cite{kawai3}.

In the case of pure 2d quantum gravity, i.e.\ $c\equ 0$ in the
terminology of non-critical string theory, $d$ measures the geodesic
distance on the underlying geometries in the path integral and
 $G_\m (x,y;d)$ can be given the following
interpretation for $x\to \infty$: It is the amplitude for a disk
where the boundary (with boundary cosmological constant $y$) is
located a geodesic distance $d$ from the ``center'' (the other
boundary contracted to a point). This amplitude is difficult to
address in Liouville theory because it is difficult to work with the
geodesic distance, which in the Liouville setup is a derived
non-local concept. However, the combinatorial  approach pioneered in
\cite{kawai1} allows a transparent derivation of eq.\ \rf{5}.
%As
%shown in \cite{aagk} eq.\ \rf{1} offers a simple interpretation of
%the transition from compact to non-compact geometry for the disk
%amplitude arising from $G_\m$:
As shown in \cite{aagk} the disk amplitude arising from $G_\m$
offers insight into the transition from compact to non-compact
worldsheet geometry. From (\ref{3}) one observes that when $d \to
\infty $ the running boundary coupling constant $\bx(d)$ converges
to the zero $x_0$ of $\hat{P}_2(x)$, which in notation of \cite{sei1} is
related to the single principal ZZ-brane in pure quantum gravity.
Moreover, for $y=-x_0$ one obtains the ``quantum'' Poincare disk
when $d \to \infty$. Hence, for this particular value of $y$ one has
a transition from a FZZT brane to a ZZ brane. Notice, this
transition is not generic: The average area and the average boundary
length of the disk remain finite in the limit $d \to \infty$ for all
other values of $y$.

For the $(2,2m\mi 1)$ minimal model coupled to 2d gravity \rf{6}
reads: \beq\label{6a} G_\m(t,t';d) \propto
\frac{1}{\sm}\;\;\frac{1}{ \bt(d) + t'}\;\;
\frac{\sqrt{1+\bt(d)}\;\prod_{n=1}^{m-1} \;(\bt(d) -
t_n)}{\sqrt{1+t}\; \prod_{n=1}^{m-1} \;(t - t_n)} \eeq where we use
the notation of \rf{1}, i.e.\ $t=x/\sm$, $t'=y/\sm$ and $\bt(d) =
\bx(d)/\sm$, where $\bx(d)$ is defined by eq.\ \rf{3}. $d$ is not
the geodesic distance for $m>2$.
%of the
%underlying geometries used in the path integral,
Rather, it is a distance measured in terms of matter excitations.
This is explicit by construction in some models of quantum gravity
with matter, for instance the Ising model and the $c\equ \mi 2$
model formulated as an $O(-2)$ model \cite{aajk,akw}. However, we
can still use $d$ as a measure of distance and we will do so in the
following. When $d \to \infty $ it follows from \rf{3} that the
running boundary coupling constant $\bt(d)$ converges to one of the
zeros of the polynomial $P_m(t)$, i.e.\ \beq\label{6b} \bt(d)
\xrightarrow[d\to \infty]{~} t_k,~~~t_k =
-\cos\left(\frac{2k\pi}{2m-1}\right). \eeq

The cylinderamplitude (\ref{6a}) vanishes for generic values of $t'$
in the limit $d \to \infty$. However, as shown in \cite{aagk} we
have a unique situation when we choose $t' = -t_k$ since in this
case the term $1/(\bt(d)+t')$ in \rf{6a} becomes singular for $d \to
\infty$. After some algebra we obtain the following expression:
\bea\label{6c} G_\m (t, t'=-t_k,d\to \infty) &\propto&
\frac{1}{\sm}\;\frac{1}{\sqrt{1+t}} \sum_{n=1}^{m-1}
(-1)^n \sin\left(\frac{2n\pi}{2m-1} \right) \\
&& \left[ \frac{1}{\sqrt{1+t}+\sqrt{1+t_n}}-
\frac{1}{\sqrt{1+t}-\sqrt{1+t_n}}\right]. \nonumber\eea Note, $G_\m
(t, t'=-t_k,d\to \infty)$ is independent of which zero $t_k$ the
running boundary coupling constant approaches in the limit $d \to
\infty$, apart from an overall constant of proportionality.

Formula \rf{6c} describes an AdS-like non-compact space with
cosmological constant $\m$ and with one compact boundary with
boundary cosmological constant $x$ as explained in \cite{aagk} in
the case of pure gravity. In the last section we will comment on the
fact that we have to set $t' = -t_k$ in order to generate an
AdS-like non-compact space in the limit $d \to \infty$. Now, we will
explain how the cylinder amplitude (\ref{6c}) is related to the
conventional FZZT--ZZ cylinder amplitude in the Liouville approach
to quantum gravity.

\section{The cylinder amplitudes}

The $(2,2m\mi 1)$ minimal CFT coupled to 2d quantum gravity has a
one-matrix representation. Using the one-matrix model one can
calculate the disk amplitude like \rf{1} or the cylinder amplitude.
Quite remarkable, the cylinder amplitude is  ``universal'', i.e.\
the same in all the $(2,2m\mi 1)$ minimal models coupled to quantum
gravity \cite{ajm,staudacher}: \beq\label{7a} C_\m (t_1,t_2) =
\frac{1}{2 \m} \frac{1}{(\sqrt{t_1+1}+\sqrt{t_2+1})^2} \;
\frac{1}{\sqrt{(t_1+1)(t_2+1)}}, \eeq where $t_1,t_2$ are boundary
cosmological constants (divided by $\sm$) and $\m$ is the bulk
cosmological constant. This amplitude is one where both
boundaries are marked, i.e.\ differentiated after the boundary
cosmological constants, and the corresponding unmarked amplitude $Z$
is: \bea\label{8} C_\m(t_1,t_2)&=&\frac{1}{\m} \frac{\prt^2}{\prt
t_1\prt t_2} Z_\m(t_1,t_2),
\\
Z_\m(t_1,t_2) &=& - \log \left[\Big(\sqrt{t_1+1}+\sqrt{t_2+1}\Big)^2
\sm \, a\right],
\label{8a}
\eea
where $a$ is a (lattice) cut-off.

The amplitude $Z_\m(t_1,t_2)$ is only one of many cylinder
amplitudes which in principle exist when we consider a $(2,2m\mi 1)$
minimal conformal field theory coupled to 2d gravity. If we consider
the cylinder amplitude of the $(2,2m\mi 1)$ minimal conformal field
theory before coupling to gravity we have available $m\mi 1$ Cardy
boundary states $|r\ra_{\mathrm{Cardy}}$, $r\equ 1,\ldots,m\mi 1$,
on each of the boundaries, and a corresponding cylinder amplitude
for each pair of Cardy boundary states \cite{bppz}:
\beq\label{9}
Z_{matter}(r,s;q) = \sqrt{2}\; b \sum_{l=1}^{m-1} (-1)^{r+s+m+l+1}
 \frac{\sin (\pi r l b^2)
\sin (\pi s l b^2)}{\sin(\pi  l b^2)}  \, \chi_l (q),
\eeq
where
\beq\label{9a} b=\sqrt{\frac{2}{2m-1}}
\eeq
and where we consider a
cylinder with a circumference of $2\pi$ and length $\pi \tau$ in the
closed string channel. The generic non-degenerate Virasoro character
$\chi_p(q)$ is \beq\label{10} \chi_p(q) =
\frac{q^{p^2}}{\eta(q)},~~~~q=\e^{-2\pi \tau}, \eeq where $\eta(q)$
is the Dedekind function. However, the degenerate Virasoro character
$\chi_l(q)$ in eq.\ \rf{9} is given by \cite{book}: \beq\label{11}
\chi_l(q)= \frac{1}{\eta(q)}\; \sum_{n \in \;\mathbb{Z}} \left(
q^{(2n/b+1/2(1/b -l\, b))^2} - q^{(2n/b+1/2(1/b+l\, b))^2}\right).
\eeq

In order to couple  the cylinder amplitude in eq.\ \rf{9} to 2d
quantum gravity one has, in the conformal gauge, to multiply
$Z_{mat}(r,s;q)$ by a contribution $Z_{ghost}(q)$ obtained by
integrating over the ghost field, as well as by a contribution
$Z_{Liouv}(t_1,t_2;q)$ obtained by integrating over the Liouville
field. Explicitly we have
\beq\label{12} Z_{ghost}(q)=
\eta^2(q),~~~~Z_{Liouv}(t_1,t_2;q)= \int_0^\infty \d P
\;\bar{\Psi}_{\sg_{1}}(P) \Psi_{\sg_{2}}(P) \chi_P(q),
\eeq
where $\Psi_\sg(P)$ is the FZZT boundary wave function \cite{fzzt}, such
that
\beq\label{13} \bar{\Psi}_{\sg_{1}}(P) \Psi_{\sg_{2}}(P) =
\frac{4 \pi^2 \cos(2\pi P \sg_{1}) \cos(2\pi P \sg_{2}) }{\sinh
(2\pi P/b ) \sinh (2\pi P b ) },
\eeq
and where $\sg$ is related to
the boundary cosmological constant by
\beq\label{14}
\frac{x}{\sm}\equiv t = \cosh (\pi b \, \sg).
\eeq
One finally obtains the full cylinder amplitude by integrating over the single
real moduli $\tau$ of the cylinder:
\beq\label{15} Z_\m
(r,t_1;s,t_2) = \int_0^\infty \d \tau \; Z_{ghost}(q)
Z_{Liouv}(t_1,t_2;q) Z_{mat}(r,s;q).
\eeq
This cylinder amplitude
depends not only on the Cardy states $r,s$, but also on the values
of the boundary cosmological constants $t_1,t_2$ as well as the bulk
cosmological constant $\m$.

From the discussion above it is natural that the matrix model (for a
specific value of $m$) only leads to a single cylinder amplitude
since it corresponds to an explicit (lattice) realization of the
conformal field theory, and thus only to one realization of boundary
conditions. In the language of Cardy states we want first to
identify {\it which} boundary condition is realized in the scaling
limits of the one-matrix model. We do that by calculating the
cylinder amplitude \rf{15} and then comparing the result with the
matrix model amplitude.

The calculation, using \rf{9}, \rf{12} and \rf{15}, is in principle
straight forward, but quite tedious. The main technical problem is
to regularize the $P$-integration around zero\footnote{One can avoid
this by working with $C(x_1,x_2)$. However, we prefer to work
directly with $Z(x_1,x_2)$ in order to compare our results with
previous calculations.} and to perform suitable deformations at
infinity. This can be done following \cite{sei2}, and the details
will be reported elsewhere \cite{ag2}. The result is for $r+s \leq
m$ \beq\label{16} Z_\m(r,t_1;s,t_2) = -
\sideset{}{'}\sum_{k=1-r}^{r-1}\;
\sideset{}{'}\sum_{l=1-s}^{s-1}\log \left(\Big[
(\sqrt{t_1+1}+\sqrt{t_2+1})^2-f_{k,l}(t_1,t_2)\Big]\sm\,a\right)
\eeq where $a$ is the cut-off (as in \rf{8}) and the summations are
in steps of two, indicated by the primes in the summation symbols.
\beq\label{17} f_{k,l}(t_1,t_2) = 4 \left[\sqrt{(t_1+1)(t_2+1)} + 2
\cos^2\left(\frac{(k+l)\pi b^2}{4}\right) \right] \; \sin^2 \left(
\frac{(k+l)\pi b^2}{4}\right). \eeq From eqs.\ \rf{16} and \rf{17}
it follows that we have agreement with the matrix model amplitude
\rf{8} if and only if $r\equ s\equ 1$. The $r\equ 1$ boundary
condition is in the concrete realizations of conformal field
theories related to the so-called fixed boundary conditions and for
the matter part of the cylinder amplitude it corresponds to the
fact, that only the conformal family of states associated with the
identity operator propagates in the open string channel.

It is worth to notice that introducing the variables
$\sg_{1},\sg_{2}$ from eq.\ \rf{14} instead of $t_1,t_2$ in eq.\
\rf{16} the cylinder amplitudes can be rewritten as: \beq\label{18}
Z_\m(r,\sg_1;s,\sg_2) = \sideset{}{'}\sum_{k=1-r}^{r-1}\;
\sideset{}{'}\sum_{l=1-s}^{s-1} Z_\m(1,\sg_1+i b k;1,\sg_2+i b l),
\eeq which shows that for $r+s \leq m$ one can express the cylinder
amplitudes $Z_\m(r;s)$ as superpositions of the $Z_\m(1;1)$
amplitudes with {\it complex} values of the boundary cosmological
constants. The composition \rf{18} of the cylinder amplitude is
fully consistent with a similar decomposition of the disk amplitude
made in \cite{sei1}. However,
%it follows from
%the vanishing of the cylinder amplitude with different pure
%Ishibashi states imposed on the  for the $(2,2m\mi 1)$ CFT that
%formula
\rf{18} is not valid for $m < r+s \le 2(m\mi 1)$. Assuming the
validity of \rf{18} for all values of $r+s \leq 2(m \mi 1)$ and of
\rf{16} for $r=s=1$, which coincides with the matrix model
amplitude, is not consistent with the vanishing of the cylinder
amplitude with two different matter Ishibashi states imposed on the
boundaries. From the fusion rules one can derive the correct
expression for $Z_\m(r,\sg_1;s,\sg_2)$ for $r+s > m$ expressed in
terms of $Z_\m(r,\sg_1;s,\sg_2)$ with $r+s \le m$, an expression
which can also be obtained by direct calculation. The explicit
expression will be reported elsewhere \cite{ag2}. Here, we only note
that any conclusion we report in this article is valid also for $r+s
> m$ unless otherwise stated.

Following Martinec \cite{martinec} it is now possible to calculate
the FZZT--ZZ amplitude by replacing one of the FZZT wave functions
in \rf{12} with \beq\label{19} \Psi_{\hn}(P)\propto
\Psi_{\sg(\hn)}(P) -\Psi_{\sg(-\hn)}(P), \eeq where (for the
$(2,2m-1$) models) \beq\label{19a} \sg(\hn) =
i\left(\frac{1}{b}+\hn\, b\right), \eeq and where $ \hn =1, \ldots,
m-1$ is an integer labeling the different principal ZZ-branes.
%For
%these complex values of $\sg(\hn)$
Notice, the boundary cosmological constants $t_\hn$ and $t_{-\hn}$
corresponding to the complex valued $\sg(\hn)$ and $\sg(-\hn)$ are
still real and are actually the same for a given value of $\hn $:
\beq\label{19b} t_{\hn} = t_{-\hn} = -\cos \Big(
\frac{2\hn\pi}{2m+1} \Big), \eeq i.e.
%according to \rf{6b}
they are the zeros of the polynomial $P_{m}(t)$ in formula \rf{1}.
%for $1 \le \hn \le m-1$
We now obtain the following FZZT--ZZ cylinder amplitude\footnote{The
upper sign in \rf{20} is for $0 \leq k+l+\hn$, while the lower sign
is for $k+l+\hn\leq 0$} for $r+s \leq m$, differentiated after the
boundary cosmological constant on the FZZT brane:
%\bea
%Z(r,\hn;s,w)&\propto & \sum_{\tm=-(r-1)}^{r-1} \sum_{\tn=-(s-1)}^{s-1}
%\left(
%\ln \left[ \sqrt{t+1} \mp
%\sqrt{1+\cos\left( \frac{2(\tm \pl\tn\mi(m\mi n))\pl 1}{2m-1}\pi\right)}\right]
%\right.\no\\
%&&-
%\left. \ln \left[ \sqrt{t+1} \pm
%\sqrt{1+\cos\left( \frac{2(\tm\pl\tn\mi(m\mi n))\pl 1}{2m-1}\pi\right)}\right]
%\right)\label{20}.
%\eea
\bea
Z_\m'(r,\hn;s,t)&\propto&
\sideset{}{'}\sum_{k=-(r-1)}^{r-1} \;\sideset{}{'}\sum_{l=-(s-1)}^{s-1}\;\;
\frac{(\pm)}{\sm \sqrt{1+t}} \label{20}\\
&&\left[ \frac{1}{\sqrt{t\pl 1} \pl \sqrt{1\pl t_{k + l + \hn}}}-
\frac{1}{\sqrt{t\pl 1} \mi \sqrt{1\pl t_{k+l + \hn}}}\right].
\nonumber \eea The differentiation after the boundary cosmological
constant is performed in order to compare with the corresponding
amplitude $G_\m(t,t'\equ -t_\hn, d\to \infty)$ given by \rf{6c},
which is the amplitude of a cylinder with one marked point on the
compact boundary.

Let us now consider the FZZT-ZZ cylinder amplitude with an $r \equ
1$ Cardy matter boundary condition imposed on the FZZT boundary.
This is the natural choice if we want to compare with the matrix
model results since the Cardy matter boundary condition captured by
the matrix model is precisely $r \equ 1$. In this case the summation
over $s$ is not present in eq.\ \rf{20} and comparing formula
\rf{20} with the expression \rf{6c} for $G_\m(t,-t_\hn,d \to\infty)$
one can show that \beq\label{21} G_\m(t,-t_\hn,d \to \infty) \propto
\sum_{r=1}^{m-1} S_{1,r} \;Z'_\m(r,\hn;1,t), \eeq where $S_{k,l} $
is the modular S-matrix in the $(2,2m\mi 1)$ minimal CFT, i.e.\
\cite{book} \beq\label{22} S_{k,l} = \sqrt{2}\, b \,
(-1)^{m+k+l}\sin (\pi kl\,b^2). \eeq This result is valid for any
$(2,2m \mi 1)$ minimal CFT coupled to quantum gravity and is valid
independent of which zero $t_{k}$ the running boundary coupling
constant approaches in the limit $d \to \infty$. The proof of
\rf{21} is straight forward but tedious and will appear elsewhere
\cite{ag2}.

The natural interpretation of eq.\ \rf{21} is that the matter
boundary state of the exit loop in the loop--loop amplitude
$G_\m(t,-t_\hn,d)$ is projected on the following linear combination
of Cardy boundary states in the limit $d \to \infty$: \beq\label{23}
|a\ra = \sum_{r=1}^{m-1} S_{1,r} \; |r\ra_{\mathrm{Cardy}} \propto
~|1\ra\ra, \eeq where the last state is the Ishibashi state
corresponding to the identity operator and where we have used the
orthogonality properties of the modular S-matrix and the relation
between Cardy states and Ishibashi states: \beq\label{24}
|r\ra_{\mathrm{Cardy}} = \sum_{k=1}^{m-1}
\dfrac{S_{r,k}}{\sqrt{S_{1,k}}} \;|k\ra\ra. \eeq

The Ishibashi state corresponding to the identity operator
is in a certain way the
simplest boundary state available,
and it is remarkable that it is precisely this
state which is captured by the explicit transition from compact to non-compact
geometry enforced by taking the distance $d \to \infty$.

\section{Discussion}

We have shown how it is possible to construct an explicit transition
from compact to non-compact geometry in the framework of 2d quantum
gravity coupled to conformal field theories. The non-compact
geometry is AdS-like in the sense that the average area and the
average length of the exit loop diverge exponentially with $d$ when
$d \to \infty$ as shown in \cite{aagk} (for pure gravity), and the
corresponding amplitude can be related to the FZZT-ZZ cylinder
amplitude with the simplest Ishibashi state living on the ZZ brane.
The $d \to \infty$ limit plays an instrumental role and we would
like to address two important aspects of this.

Firstly, in \cite{sei1} it was advocated that the algebraic surface
\beq\label{25} T_p(w/C_{p,q}(\mu)) =T_q(t), \eeq where $C_{p,q}(\m)$
is a constant, is the natural "target space" of $(p,q)$ non-critical
string theory. For $(p,q)=(2,2m \mi 1)$ eq.\ \rf{25} reads
\beq\label{25a} w^2 = \mu^{\frac{2m-1}{2}}P_m^2(t)(t+1), \eeq and in
this case the extended target space is a double sheeted cover of the
complex $t$-plane except at the singular points, which are precisely
the points $(t_k,w\equ 0)$ associated with the zeros of the
polynomial $P_m(t)$. One is also led to this extended target space
from the world-sheet considerations made here. We want the running
boundary coupling constant to be able to approach any of the fixed
points $t_k$ in the limit $d \to \infty$, i.e.\ we want all the
fixed points to be attractive. This is only possible if we consider
the running boundary coupling constant $\bt(d)= \bx(d)/\sm$ defined
in eq.\ \rf{3} as a function taking values on the algebraic surface
defined by \rf{25a}. The reason is that $t_k$ is either an
attractive or a repulsive fixed point depending on which sheet we
consider and some of the fixed points are attractive on one sheet,
while the other fixed points are attractive on the other sheet.
%from which it is approached.
%We only allow the distance $d$ to take on
%positive values. For odd values of $m$ and $t > t_{m-1}$ the running
%boundary coupling constant $\bt(d)= \bx(d)/\sm$ (defined in eq.\
%\rf{3}) approaches infinity on the first sheet for a finite positive
%value of $d$ and then returns on the second sheet as we increase $d$
%even further.
%want $d$ to be positive and
%go to infinity.
%However, this is only possible if we consider
Hence, we are forced to view $\bt(d)$ as a map to the double sheeted
Riemann surface defined by eq.\ \rf{25a} in the $(2,2m \mi 1)$
minimal model coupled to quantum gravity.

The picture becomes particularly transparent if we use the
uniformization variable $z$ introduced for the $(p,q)$ non-critical
string in \cite{sei1} by \beq\label{26}
t=T_p(z),~~w/C_{p,q}(\m)=T_q(z), \eeq i.e.\ in the case of
$(p,q)=(2,2m \mi 1)$: \beq\label{26a} z=\frac{1}{\sqrt{2}}
\sqrt{t+1}. \eeq The map \rf{26} is one-to-one from the complex
plane to the algebraic surface \rf{25}, except at the singular
points of the surface where it is two-to-one. The singular points
are precisely the points corresponding to ZZ branes.
%In the
%case of the $(2,2m \mi 1)$ theories the singular points are
%precisely the $(t_k,w\equ 0)$ associated with the zeros of the
%polynomial $P_m(t)$.
If we change variables from $x$ to $z$ in eq.\
\rf{5} (choosing $\m \equ 1$ for simplicity) we obtain
\beq\label{27}
 \frac{\prt }{\prt d}\; \tG_\m (z,z';d) = - \frac{\prt }{\prt z} \;
\tP_m(z) \tG_\m (z,z';d), \eeq where $\tG_\m(z,z';d)=zG_\m(x,y;d)$
and where the polynomial $\tP_{m}(z)$ is \beq\label{28} \tP_{m}(z)
\propto \prod_{k=1}^{m-1} (z^2 -z_k^2), ~~~~z_k = \sin
\left(\frac{\pi}{2} \, b^2 \, k\right). \eeq Each zero $t_k$ of
$P_m(t)$ gives rise to two zeros $\pm z_k$ of $\tP_m(z)$.
%Like the $t_k$'s
The zeros $\pm z_{k}$ are the fixed points of the running
``uniformized'' boundary cosmological constant $\bar{z}$ associated
with the characteristic equation corresponding to eq.\ \rf{27}. For
a given value of $k$ one of the two zeros $\pm z_k$ is an attractive
fixed point, while the other is repulsive. Moving from one sheet to
the other sheet on the algebraic surface (\ref{25a}) corresponds to
crossing the imaginary axis in the $z$-plane. Hence, for a given
value of $k$ the two fixed points $\pm z_{k}$ are each associated
with a separate sheet and $\bar{z}$ will only approach the
attractive of the two fixed points $\pm z_{k}$, if $\bt(d)$ belongs
to the correct sheet.
%and choosing the correct sheet of the
%Riemann surface \rf{25} for $\bt(d)$ corresponding to $d\to \infty$
%corresponds to approaching the attractive of the two fixed points
%$\pm z_k$.

Quite remarkable eq.\ \rf{27} was derived in the case of pure 2d
gravity (the $(2,3)$ model corresponding to $c\equ 0$) using a
completely different approach to quantum gravity called
CDT\footnote{It should be noted that CDT seemingly has an
interesting generalization to higher dimensional quantum gravity
theories \cite{ajl1,ajl2}} (causal dynamical triangulations)
\cite{al} and the uniformization transformation relating the CDT
boundary cosmological constant $z$ to the boundary cosmological
constant $t$ was derived and given a world-sheet interpretation in
\cite{ackl}, but again from a different perspective. From the CDT
loop-loop amplitude determined by \rf{27} one can define a CDT ``ZZ
brane'' with non-compact geometry \cite{ajwz}.

Secondly, our construction also adds to the understanding of the
relation \rf{19} discovered by Martinec. In Liouville theory there
is a one-to-one correspondance between the ZZ boundary states
labeled by $(m,n)$ and the degenerate primary operators $V_{m,n}$
\cite{zz}. This correspondance completely determines the Liouville
cylinder amplitude with two ZZ boundary conditions: The spectrum of
states flowing in the open string channel between two ZZ boundary
states is obtained from the fusion algebra of the corresponding
degenerate operators. Similarly, there is a one-to-one
correspondance between the FZZT boundary states labeled by $\sg>0$
and the non-local "normalizable" primary operators
$V_{\sg}=\exp((Q+i\sg)\phi)$, where $\phi$ is the Liouville field.
The conformal dimension of the spin-less degenerate primary operator
$V_{m,n}$ is given by \beq\label{29}
\Delta_{m,n}=\frac{Q^{2}-(m/b+nb)^{2}}{4}, \eeq while the conformal
dimension of the spin-less non-local primary operator $V_{\sg}$ is
given by \beq\label{30} \Delta_{\sg}=\frac{Q^{2}+\sg^{2}}{4}. \eeq
Since $\Delta_{m,n}=\Delta_{\sg}$ for $\sg=i(m/b+nb)$, one is
naively led to the wrong conclusion, that a FZZT boundary state
turns into a ZZ boundary state, if one tunes $\sg=i(m/b+nb)$.
However, the operator $V_{m,n}$ is degenerate and in addition to
setting $\sg=i(m/b+nb)$ we therefore have to truncate the spectrum
of open string states, that couple to the FZZT boundary state, in
order to obtain a ZZ boundary state. This is precisely captured in
the relation \rf{19} concerning the principal ZZ boundary states.
The world-sheet geometry characterizing the FZZT brane is compact,
while the world-sheet geometry of the ZZ-brane is non-compact.
Hence, truncating the spectrum of open string states induces a
transition from compact to non-compact geometry. In our concrete
realization of this transition this truncation is obtained by first
setting the boundary cosmological constant $t'=-t_{\hn}$ on the exit
loop and then taking $d\to \infty$. It is interesting to note that
in the original articles introducing the FZZT and ZZ boundary states
\cite{fzzt,zz} the square of eq.\ \rf{14} is always used as the
defining relation between $\sg$ and $t$. Thus, in these works both
$\pm t_{\hn}$ are associated with the ZZ brane labeled by $(1,\hn)$
through eq. (\ref{19}). In our explicit construction both values
also play a role: $t_\hn$ is the fixed point of the running boundary
coupling constant $\bt(d)$ as $d \to \infty$ and $- t_{\hn}$ is the
actual value (measured in units of $\sqrt{\mu}$) of the boundary
cosmological constant on the "AdS-boundary". Only for this
particular value of the boundary cosmological constant does an AdS
geometry emerge
%where the area and the length of the circumference diverge
in the limit $d \to \infty$.

\section*{Acknowledgment}
Both authors acknowledge the  support by
ENRAGE (European Network on
Random Geometry), a Marie Curie Research Training Network in the
European Community's Sixth Framework Programme, network contract
MRTN-CT-2004-005616.

\end{document}